\documentclass[12pt,preprint]{aastex}

\def\hii{H\,{\sc ii}}

\def\kms{\relax \ifmmode {\,\rm km~s}^{-1}\else \,km~s$^{-1}$\fi}
\def\cm-3{\relax \ifmmode {\,\rm cm}^{-3}\else \,cm$^{-3}$\fi}
\def\Jb{\relax \ifmmode {\,\rm Jy\,beam}^{-1}\else \,Jy\,beam$^{-1}$\fi}
\def\mJb{\relax \ifmmode {\,\rm mJy\,beam}^{-1}\else \,mJy\,beam$^{-1}$\fi}
\def\deg{\hbox{$^\circ$}}
\def\arcmin{\hbox{$^\prime$}}
\def\arcsec{\hbox{$^{\prime\prime}$}}
\def\secd#1.#2{ #1\farcs#2 }               
\def\e{$\pm$}
\def\x{$\times$}
\def\j21{$J$=2$\rightarrow$1}

\begin{document}

\title{Interferometric Mapping of Magnetic Fields in Star-forming Regions III.  Dust and CO polarization in DR21(OH)}
\author{Shih-Ping Lai\altaffilmark{1}, Jos\'e M. Girart\altaffilmark{2}, and Richard M. Crutcher}
\affil{Astronomy Department, University of Illinois, 1002 W. Green Street, Urbana, IL 61801; \\
slai@astro.umd.edu, jgirart@am.ub.es, crutcher@astro.uiuc.edu}
\altaffiltext{1}{Current address: Department of Astronomy, University of Maryland, College Park, MD 20742-2421}
\altaffiltext{2}{Current address: Departament d'Astronomia i Meteorologia, Universitat de Barcelona, 08028 Barcelona, Catalunya, Spain}

\begin{abstract}
We present the polarization detections in DR21(OH) from both the thermal
dust emission at 1.3 mm and the CO \j21\ line 
obtained with the Berkeley-Illinois-Maryland Association (BIMA) array.  
Our results are consistent with 
the prediction of the Goldreich-Kylafis effect that the CO polarization 
is either parallel or perpendicular to the magnetic field direction.
The detection of the polarized CO emission is over a more extended
region than the dust polarization,
while the dust polarization provides an aide in resolving 
the ambiguity of the CO polarization.
The combined results suggest that the magnetic field direction in DR21(OH)
is parallel to the CO polarization and therefore 
parallel to the major axis of DR21(OH).
The strong correlation between the CO and dust polarization
suggests that magnetic fields are remarkably uniform throughout 
the envelope and the cores.
The dispersion in polarization position angles implies a magnetic 
field strength in the plane of the sky of about 1 mG,
compared with about 0.5 mG inferred for the line-of-sight field 
from previous CN Zeeman observations.
Our CO data also show that both MM1 and MM2 power high-velocity 
outflows with $v\gtrsim$25 \kms\ relative to the systematic velocity.

\end{abstract}

\section{Introduction}

In order to study the importance of the magnetic field in the star
formation process, it is essential to obtain high resolution maps
of magnetic fields in dense molecular cores where stars form.
Mapping polarized thermal dust emission has been the most successful
technique to explore the magnetic field morphology in dense cores
(Greaves et al.\ 1999a; Dotson et al.\ 2000; Ward-Thompson et al.\ 2000;
Matthews \& Wilson 2002).
In general, the magnetic field direction is perpendicular to the 
dust polarization direction.  Using the BIMA millimeter interferometer, 
we have obtained high resolution maps of dust polarization 
in several star-forming cores (Rao et al.\ 1998; Girart, Crutcher, \& Rao 
1999; 
Lai et al.\ 2001, hereafter Paper I; Lai et al.\ 2002, hereafter Paper II).
Our results show that the magnetic fields in these cores are remarkably 
uniform, suggesting that magnetic fields are relatively strong and 
cannot be ignored in cloud evolution and star formation.

Polarization arising from the spectral line emission, 
via the Goldreich-Kylafis effect, 
provides an opportunity to probe the magnetic field directions in 
the regions where the dust emission is too weak for polarization detections.
This effect predicts that the line polarization has a direction
either perpendicular or parallel to the magnetic field direction,
depending on the relative angles between the magnetic field, the velocity 
gradient, and the line of sight (Goldreich \& Kylafis 1981, 1982; Kylafis 1983).
Detection of this effect requires high sensitivity and spatial
resolution in order to separate regions with different physical conditions;
therefore, the observational confirmation of this effect comes very recently
(Greaves et al.\ 1999b; Girart, Crutcher, \& Rao 1999).
Here we present another example, DR21(OH), with polarization
detections in both the dust emission and in the CO \j21\ line.
Simultaneous observations of both effects help us to extend the mapping 
area of the magnetic field and to resolve the ambiguity of 
the predicted field direction from the CO polarization.

DR21(OH), also known as W75S or W75S-OH, is located 3\arcmin\ north of 
the \hii\ region DR21 in the Cygnus X molecular cloud/\hii\ region complex
(Harvey et al.\ 1986; Downes \& Rinehart 1966).
The distance to DR21(OH) is commonly assumed to be 3 kpc,
although the value is very uncertain (Campbell et al.\ 1982).
Its association with masers of OH (Norris et al.\ 1982),
H$_2$O (Genzel \& Downes 1977), and CH$_3$OH (Batrla \& Menten 1988;
Plambeck \& Menten 1990) suggests the presence of high-mass,
young stellar objects.  
The main component of DR21(OH) has been resolved into two compact cores, 
MM1 and MM2 (Woody et al.\ 1989), with a total mass of $\sim$ 100 M$_\sun$ 
in roughly virial equilibrium (Padin et al.\ 1989).   
The Zeeman splitting in CN lines, which trace high density 
of $n_{H_2}\sim10^5-10^6$\cm-3, has been detected in both cores.
In fact, MM1 and MM2 are two of the three cores in which 
the Zeeman effect in CN lines has been detected to date.
The line-of-sight magnetic field strength, $B_{los}$, was determined
to be $-$0.4 mG for MM1 and $-$0.7 mG for MM2 (Crutcher et al.\ 1999).
Therefore, by mapping magnetic field directions in the plane of sky 
from dust and CO polarization, we have an opportunity
to explore the magnetic field of DR21(OH) in 3-dimensional space.

\section{Observations and Data Reduction}

The observations were carried out from 1999 December to 2000 May
using nine BIMA antennas with 1-mm Superconductor-Insulator-Superconductor
(SIS) receivers and quarter-wave plates.
The digital correlator was set up to observe both continuum and the
CO \j21\ line simultaneously.  The continuum was observed
with a 750 MHz window centered at 226.9 GHz in the lower sideband
and a 700 MHz window centered at 230.9 GHz in the upper sideband.
An additional 50 MHz window in the upper sideband was set up to
observe the CO $J$=2--1 line.
The primary beam was $\sim$50\arcsec\ at 1.3 mm wavelength.
Data were obtained in the C and D array configurations,
and the projected baseline ranges were 5--68 and 4.5--20 kilowavelengths.
The on-source integration time in the C and D array was 23.1 and 2.7 hours,
respectively.

The BIMA polarimeter and the calibration procedure are described
in detail in Paper I.  The average instrumental polarization of each antenna,
which was removed from the DR21(OH) data by our standard calibration
procedure, was 6.2\%$\pm$0.5\% for our observations.  
The dust continuum images 
of Stokes $I, Q$, and $U$ were  made with natural weighting and 
the resulting synthesized beam is 3\farcs9\x3\farcs5 (PA=12\deg).
The CO images were made with a Gaussian taper applied to the CO visibilities 
in order to increase the sensitivity per beam, and the synthesized 
beam for CO is 5\farcs7\x4\farcs6 with PA=$-$7\deg.
These maps were binned to approximately half-beamwidth per pixel 
(2\farcs0\x1\farcs6) to reduce oversampling in our statistics.
The binned maps were then combined to obtain
the linearly polarized intensity ($I_p$), the position
angle ($\phi$), and the polarization percentage ($p$),
along with their uncertainties as described in Section 2 of Paper I.
When weighted with $I_p$, the average measurement uncertainty in
the position angle for our observations was 7\fdg6\e1\fdg2.

\section{Results}

Figure 1 shows the dust continuum in contours and the CO \j21\ 
emission ($v_{LSR}$ of $-$10 \kms, $\Delta v$=11.5 \kms) 
in grey scale overlaid with polarization vectors of dust and CO.   
The CO emission in Figure 1 is only integrated over the velocity 
range associated with the polarization detections.
The dust polarization vectors are plotted at positions
where the linearly polarized intensity $I_p$ is greater than
3$\sigma_{I_p}$
(1$\sigma_{I_p}$ = 2.0 \mJb) and the total intensity $I$
is greater than 3$\sigma_{I}$ 
(1$\sigma_{I}$ = 4.0 \mJb).
The CO polarization vectors are shown at positions
with the linearly polarized intensity greater than
3$\sigma_{I_p}$ (1$\sigma_{I_p}$ = 29 \mJb) and the integrated
intensity greater than 30$\sigma_{I}$ (1$\sigma_{I}$ = 0.13 \Jb).
The total area with polarization detections is $\sim$ 6 beam sizes
for dust and $\sim$ 7 beam sizes for CO.
Table 1 lists the dust and CO polarization vectors shown in Figure 1.
The separation between vectors is approximately half of the synthesized beam.

\subsection{Dust Emission and Polarization}

DR21(OH) is comprised of two continuum sources,
MM1 in the east and MM2 in the west (Fig.\ 1).
Both sources are resolved, and their sizes are 6\farcs5\x5\farcs2 
(PA=30\deg) for MM1 and 7\farcs4\x6\farcs6 (PA=15\deg) for MM2.
The physical sizes are $\sim$ 0.1 pc at a distance of 3 kpc.
MM1 and MM2 each has a flux of 1.6 Jy.
Since the free-free emission in these two cores is negligible
(Johnston, Henkle, \& Wilson 1984),
the continuum flux of DR21(OH) at 1.3 mm is dominated by the 
dust emission.

The polarization detections in the dust continuum of DR21(OH) appear 
in several scattered positions.
Figure 2(a) shows that the distribution of the polarization 
angles in the dust continuum of DR21(OH); 
the average angle is $-$38\deg\ and the dispersion is 28\deg\e9\deg.
This large dispersion includes the dispersion caused by the turbulent 
motions as well as the variation of the magnetic field structure across
the double cores.    For further examination, we divide 
the polarization detections into three groups according to their spatial 
locations, which are near MM1, MM2, and the northern edge of DR21(OH).
The histograms of the polarization angles in these three groups
are shown separately in Figure 2(b)-(d).
The polarization emission associated with MM1 appears in two 
sub-groups with distinct polarization angles:
one in the northeast side of the MM1 peak with polarization 
angles at $-$10\deg\e2\deg, and the other in the southern part of MM1
with polarization angles at $-$50\deg\e7\deg.
The vectors associated with MM2 show more continuous change between
$-$35\deg\ to 15\deg\ with an average of $-18\deg\pm12\deg$.  
The vectors in the northern edge have position angles at $-$86\deg\e12\deg, 
which are very different from those associated with
MM1 and MM2.   We discuss the four possibilities for the difference
between the northern edge and the two main cores in \S4.3.
Overall, although we cannot completely decouple the variation 
of the magnetic field structure from the angle dispersion,
we may use the dispersion of MM1 and MM2 (21\deg\e7\deg) 
as the upper limit of the angle dispersion associated with
the turbulent motions in the dust cores.

The distributions of the polarization percentage
versus the total intensity and the distance to the peaks of MM1 and MM2
are plotted in Figure 3.
Fig.\ 3(a) shows that the polarization percentage decreases toward
regions of high intensity.
The least-squares fit on $\log_{10}p$ versus $\log_{10}I$ 
for all data points gives $\log_{10}p=(-2.08\pm0.01)-(0.95\pm0.01)
\times\log_{10}I$ with a correlation coefficient of $-0.98$.
Fig.\ 3(b) and (c) show that
the polarization percentage decreases toward the center of both cores.
Because higher intensity and smaller radius both imply higher density,
our results suggest that the polarization percentage decreases toward
high density regions.    This is consistent with what we have observed in 
W51 e1/e2 and NGC 2024 FIR 5 (Paper I and II),
and we again attribute this to the decrease of the dust 
polarization efficiency toward high density regions. 

\subsection{CO \j21\ Emission and Polarization}

Figure 4 shows the Stokes $I, Q$, and $U$ spectra of CO \j21\ emission 
integrated over the region associated with the polarization peak
in the center of the dust continuum (the region is shown with
a dashed rectangle in Figure 1).
The CO line is clearly optically thick and
contains emission from both the dense core and the ambient gas,
resulting in a wide linewidth in Stokes $I$.
The Stokes $I$ line profile shows two peaks at $v_{LSR}\sim-10$ and 0 \kms\
with a minimum at $-$4\kms.
The velocities of these peaks are different from previous CS and CN 
observations, which are at $v_{LSR}\sim-5$ and $-1$ \kms\
(Richardson et al.\ 1994; Crutcher et al.\ 1999).
Therefore, the dip in the Stokes $I$ line profile should be 
due to self-absorption and/or severe missing flux.

The CO polarized flux is mostly associated with 
the $-$10 \kms\ peak in the spectrum (Fig.\ 4).
The polarization map of this peak is shown in Fig.\ 1.
The detection of the polarized CO emission is over a more extended
region than the dust polarization.
Most of the polarized emission arises from a region associated 
with the dust continuum.  We refer to this region as CO pol Main.
The polarization angles in CO pol Main appear to decrease from
east to west from 140\deg\ to 40\deg\ with an average of 89\deg\e23\deg.
There is also a polarized region with an area $\sim$1--2 beamsizes 
at $\sim$15\arcsec\ northwest of the MM2 peak, which we refer
to CO pol West.  The average polarization angle in CO pol West 
is 59\deg\e10\deg.  The polarized flux drops to zero between
CO pol Main and CO pol West, suggesting these is significant change
in the field geometry between these two regions.

\subsection{Comparison between Dust and CO Polarization}

Figure 5 shows the distribution of the CO polarization angles
in CO pol Main
and Figure 6 shows the distribution of the difference between 
the polarization angles of CO and dust ($\phi_{CO}-\phi_{dust}$)
for those positions in which detections have been made in both
CO and dust emission.  The Goldreich-Kylafis effect predicts that the CO 
polarization direction is either parallel or perpendicular to the 
magnetic field direction.  Therefore, 
in theory $\phi_{CO}-\phi_{dust}$ should be either 0\deg\ or 90\deg.   
Our results are consistent with the theoretical predictions:
near the northern edge, $\phi_{CO}$ is approximately parallel to 
$\phi_{dust}$; near the MM1 and MM2 cores, the CO polarization 
vectors are close to perpendicular to the dust polarization vectors
with average $\phi_{CO}-\phi_{dust}$=95\deg and a dispersion 
of 31\deg\e16\deg.
If there is no correlation between the CO and dust polarization,
the expected $\phi_{CO}-\phi_{dust}$ distribution would be flat 
because the chance for any angle difference are equal.
The large dispersion could be explained by the combination 
of the intrinsic dispersion in dust and CO polarization (see Table 2),
which is 30\deg\e8\deg.  Alternatively, it could be due to the fact
that CO and dust do not trace exactly the same region, as the CO is 
optically thick and the dust emission is optically thin at 1mm.
The strong correlation between the CO and dust polarization
suggests that magnetic fields are remarkably uniform throughout 
the envelope and the cores.

\subsection{High-velocity CO outflows}

The existence of the high velocity gas in DR21(OH) has been suggested 
by Fischer et al.\ (1985) and Richardson et al.\ (1994) 
with high-velocity wings in the CO and CS line profiles.  
Our CO data provide the first map of the spectacular outflows in DR21(OH)
(Fig.\ 7).
The blue and red lobes shown in Fig.\ 7 are obtained by integrating over
7\kms\ bandwidths centered at $-$30\kms\ and 20\kms, which are
near the two ends of our correlator window.
There are other possible outflow features at lower velocities; however,
the CO emission suffers from the missing flux problem.
The outflows could extend to higher velocities, as the intensity
in the two end channels are still well above the rms noise (Fig.\ 4).
Therefore, Fig.\ 7 only presents a partial picture of the outflows in DR21(OH).

Nevertheless, from the morphologies of the CO gas with $v \la 25$~km~s$^{-1}$
relative to the systematic velocity, it seems that MM1 and MM2 each power
high-velocity bipolar outflows.  The northwestern blue lobe and the
southeastern red lobe likely originate from MM2.  Shock excited methanol
masers appear aligned with these two lobes and MM2 (Plambeck \& Menten 1990).
The southwestern blue lobe and the northeastern red lobe could originate
from both MM1 and MM2. 
In this picture the overlapped red and blue emission between MM1 and
MM2 arises from the two outflow sources.  
Alternatively, the morphology could suggest
a single bipolar outflow with a cone-like morphology, and with the CO lobes
tracing the limb brightened region of the outflow. 
In principle, this hypothesis can be examined with the position-velocity 
diagram along the jet axis (Lee et al.\ 2000);
however, the combination of the lack of extended emission and
the possible complexity of the sources impede the diagnostic.
MM1 and MM2 are both massive enough to contain multiple young stellar
objects.  In fact, MM2 has been resolved into two NH$_3$ cores with 
VLA observations (Mangum, Wootten, \& Mundy 1992).
On the other hand, the fact that the maser spots are either in or near
the outflow lobes (Fig.\ 7), but not between the lobes,
seems to favor the idea that these lobes are individual outflows.
Further single-dish observations as well as higher resolution observations 
would help to understand the driving sources and the kinematics of
the outflows in DR21(OH).

\section{Discussion}

\subsection{Magnetic Field Morphology}

We attempt to construct the magnetic field morphology in DR21(OH)
using both dust and CO polarization results. 
The dust polarization is perpendicular to the average 
field direction along the line of sight,
if produced by magnetic alignment.  The CO polarization angles 
could be either parallel or perpendicular to the field, according
to the Goldreich-Kylafis effect.

Our CO polarization map in DR21(OH) shows that CO polarization
provides measurements for the field structure in a more extended 
region than the dust polarization.
Our dust polarization results suggest that the magnetic field directions 
are parallel to the CO polarization vectors in the south of the
MM1 and MM2 peaks. But in the region to the northern edge 
where dust polarization is detected, 
the position angle of dust polarization has changed by about 90\deg.
This could suggest that the magnetic field direction is perpendicular 
to the CO polarization in the northern region; however,
it seems unlikely that the CO polarization traces such 
an abrupt change in the field direction, but still shows smooth
variation in polarization angles from south to north.
Although the dust polarization detections in the northern edge
are marginal (Table 1), they have polarization angles perfectly
consistent with the CO polarization, suggesting that they are 
probably not spurious.  
The unusual behavior of the dust polarization
in the northern portion of the cores will be discussed in \S4.3.
Here we conclude that the field morphology in DR21(OH)
is better represented by the CO polarization directions,
which is approximately along the major axis of DR21(OH).
Therefore, MM1 and MM2 are likely to be two condensations
in a magnetic flux tube.

\subsection{Magnetic Field Strengths and Directions}

We use our measurements of polarization angle 
dispersion and the Zeeman measurements from Crutcher et al.\ (1999)
to estimate the magnetic field strength and direction of DR21(OH).
As discussed in Paper I and Paper II, the Chandrasekhar-Fermi (CF) 
equation with a correction factor of 0.5 may be used to estimate
the plane-of-sky magnetic field strength ($B_p$).
The measurements of the polarization angle dispersion associated 
with Alfv\'enic motion ($\delta\phi$), the dispersion of the turbulent
linewidth, and the average density of the core are needed
for this estimate.  
There are two effects that could bias the measured $\delta\phi$ 
from its actual value:
First, the contribution due to bending of the uniform magnetic field 
has not been taken out, which would increase $\delta\phi$.
Second, any smoothing of the polarization morphology due to
inadequate angular resolution would reduce $\delta\phi$.
Since it is difficult to exclude these two factors in $\delta\phi$,
we simply use $\delta\phi\sim22$\deg\ of CO pol Main as an estimate 
(Table 2).
The linewidth and density of DR21(OH) are adopted from Crutcher et al.\ 
(1999): the FWHM of CN is 2.3 \kms\ toward both cores and
the density is 10$^6$ \cm-3\ for MM1 and 2\x10$^6$ \cm-3\ for MM2.  
Therefore, the derived $B_p$ is $\sim$0.9 mG for MM1 and $\sim$1.3 mG 
for MM2.  It is interesting that the values for $B_p$ are about 
twice the line-of-sight strengths measured with the CN Zeeman effect
($B_{los}=-$0.4 mG for MM1 and $-$0.7 mG for MM2).
Combining the field strengths in the plane of sky and in the line of 
sight, we find that the magnetic fields in MM1 and MM2 are 
pointed toward us and have an angle $\sim$ 30\deg\ 
to the plane of the sky.  
Although the CO dispersion is probably not probing the same part of
the magnetic field as the CN Zeeman data, 
our estimate of the total field strength could be useful
because the field morphology seems to be fairly uniform 
throughout the envelope and the cores (\S 3.3).

\subsection{Apparent Anomalies in Dust Polarization}

Our dust polarization detections are patchy in DR21(OH).   
As discussed in \S4.1, if the polarization in the northern patch 
follows the prediction for the magnetic alignment, it will be difficult
to construct a smooth magnetic field geometry.
There is also a large polarization gap
northwest of the peaks of the double cores
and south of the polarized region in the northern edge
of the cores (Fig.\ 1).  
The relation between the polarization percentage and
the total intensity derived in \S3.1 and shown in Fig.\ 3
fails in this polarization gap,
suggesting that there are some physical parameters
controlling the degree of polarization other than
the magnetic alignment efficiency of the dust grains.

We discuss several possible solutions for these apparent anomalies
in dust polarization.
(1) A twisted field structure could cause low polarization;
however, the smooth distribution of the CO polarization 
contradicts this explanation.
(2) If the magnetic field direction is almost parallel 
to the line of sight, slight variation in the field direction
can result in the different angles for polarizations;
however, this is not consistent with the magnetic field direction 
we derived in \S4.2 unless the CF formula is inadequate for
estimates of the field strengths.
(3) It is possible that the dust polarization in the northern edge originates 
from dust alignment mechanisms that can produce polarization angles 
parallel to the field directions,
such as the Gold alignment caused by strong gas streams (Gold 1952;
Lazarian 1994, 1997).
The change of the alignment mechanism has been suggested 
to be responsible for the abrupt polarization angles changes
in the Orion-KL region (Rao et al.\ 1998).
If there are unseen outflows (i.e. that show up at radial velocities lower
than 25~km~s$^{-1}$ relative to the systematic velocity),
the combination of the mechanical alignment
and the magnetic alignment of the dust grains 
could produce the polarization gap as
these two effects generate orthogonal polarization.    
(4) The dust polarization detections at the northern edge could 
originate from an additional source along the line of sight.
The gap can therefore be produced by the cancellation of orthogonal 
polarization.
More sensitive dust polarization data and kinematic information are 
needed to examine these four possibilities.

\subsection{Outflows and Magnetic fields}

In Figure 7, we overlaid the CO outflows on the inferred magnetic field
directions from CO and dust polarizations (assuming the dust is
aligned by magnetic fields).
The outflows seem to follow the field directions near MM1 and MM2, but depart from the field directions away from the two cores.
These results seem to be contrary the theoretical expectations
that the outflows are collimated or bent by magnetic fields
(Shu et al.\ 1995; Hurka, Schmid-Burgk, \& Hardee 1999).
However, the scales at which the outflow are collimated, $\sim$ 1 AU 
(Shu et al.\ 1995), are much smaller than that traced with the BIMA 
angular resolution. In addition, 
the CO polarization and the high-velocity outflows
are detected in different velocity ranges; 
the CO polarization detection is associated with the $-10$\kms\ peak,
which is between the ambient velocity and the velocity of the blue lobe.
It is possible that the outflows have escaped from the cores, 
so the field direction in the cores cannot be used
to represent that in the outflows.   A proper comparison between
the outflows and the magnetic field is better made with 
the detection of CO polarization in the outflow velocity range.

\section{Conclusion}

We have measured linear polarization of the thermal dust emission
at $\lambda\sim$1.3 mm and the CO \j21\ emission toward DR21(OH).

\begin{itemize}
\item The CO polarization is approximately perpendicular to 
the dust polarization associated with MM1 and MM2 and 
parallel to the dust polarization in the northern edge of the cores,
which is consistent with the predictions of the Goldreich-Kylafis
effect.

\item The extended CO polarization of DR21(OH) provides a better presentation
of the field morphology in DR21(OH) in the plane of the sky, 
which is approximately parallel to the long axis of the double cores.  

\item The plane-of-sky magnetic field strength $B_p$ estimated from
the Chandrasekhar-Fermi formula is $\sim$ 0.9 mG for MM1 and $\sim$ 
1.3 mG for MM2.   Combined with the previous CN Zeeman measurements,
we find the magnetic field directions of MM1 and MM2 
in 3-D space are both pointed toward us and have an angle
$\sim$ 30\deg\ to the plane of the sky.

\item Our CO data provide the first maps of the high-velocity outflows
originating from MM1 and MM2.  The outflow direction in the downstream
disagrees with the field direction of the cores, suggesting that
the outflows have escaped the cores to the extend that the core magnetic
field is not relevant to the outflows.

\end{itemize}

\acknowledgements
This research was supported by NSF grants AST 99-81363 and 02-05810. 
J.\ M.\ G.\ acknowledges support by MCyT grant AYA-2002-00205.

\begin{table}
\begin{center}
\caption{Polarization measurements in DR21(OH)}
\vspace*{0.3cm}
{\tiny
\begin{tabular}{ccccccl}
\hline
\hline
			& \multicolumn{2}{c}{Dust} && \multicolumn{2}{c}{CO} & \\
\cline{2-3}\cline{5-6}
Offsets\tablenotemark{a} & p & $\phi$  & & p & $\phi$ & Note \\
	 (\arcsec,\arcsec) & (\%) & (\deg) && (\%) & (\deg)&\\
\hline
 (-2.2,-6.0) & ~9.0\e ~3.2 & -17\e10 & &           & 		 &\\
 (-3.8,-6.0) & 12.1\e ~3.9 & -10\e~9 & &         &		 &        \\
 (~5.8,-4.0) & 12.9\e ~3.6 & -59\e~8 & &         &		 &        \\	
 (-0.6,-4.0) &               &       & & 0.7\e0.2&   ~50\e~9 & \\
 (-2.2,-4.0) & ~4.0\e ~1.1 & -27\e~8 & & 0.9\e0.2&   ~50\e~8 & \\		
 (-3.8,-4.0) & ~6.8\e ~1.3 & -23\e~6 & & 0.7\e0.2&   ~48\e~9 & polarization peak near MM2 \\
 (-5.4,-4.0) & ~7.1\e ~2.2 & -16\e~9 & &         &		 &        \\
 (12.2,-2.0) &               &       & & 1.2\e0.4&   -86\e10 &        \\
 (~7.4,-2.0) & ~7.5\e ~2.6 & -40\e10 & &         &		 &        \\
 (~5.8,-2.0) & ~6.6\e ~1.7 & -54\e~7 & &         &		 &        \\
 (~4.2,-2.0) & ~6.1\e ~1.4 & -48\e~6 & &         &		 & \\ 
 (~2.6,-2.0) & ~6.3\e ~1.3 & -47\e~6 & &      	      &		 & polarization peak near MM1   \\
 (~1.0,-2.0) &             &         & & 0.6\e0.2&   ~61\e~7 &        \\
 (-0.6,-2.0) & ~2.9\e ~0.8 & -30\e~8 & & 0.9\e0.2&   ~60\e~5 &        \\
 (-2.2,-2.0) & ~3.0\e ~0.6 & -29\e~6 & & 0.9\e0.2&   ~57\e~5 &        \\
 (-3.8,-2.0) & ~3.5\e ~0.7 & -23\e~6 & & 0.8\e0.2&   ~58\e~6 &        \\
 (-5.4,-2.0) & ~4.4\e ~1.2 & -~7\e~8 & & 0.6\e0.2&   ~69\e~8 &        \\
 (12.2,~0.0) &               &       & & 1.0\e0.3&   -70\e~8 &        \\
 (10.6,~0.0) &               &       & & 0.7\e0.3&   -75\e10 &        \\
 (~5.8,~0.0) &               &       & & 0.7\e0.2&   ~84\e~8 &        \\
 (~4.2,~0.0) & ~2.6\e ~0.7 & -50\e~8 & & 0.7\e0.2&   ~85\e~7 & \\
 (~2.6,~0.0) & ~2.8\e ~0.7 & -51\e~7 & & 0.6\e0.1&   ~80\e~7 &        \\
 (~1.0,~0.0) & ~2.5\e ~0.9 & -33\e10 & & 0.7\e0.1&   ~71\e~5 &        \\
 (-0.6,~0.0) & ~2.8\e ~0.7 & -20\e~7 & & 1.0\e0.1&   ~69\e~4 &        \\
 (-2.2,~0.0) & ~2.1\e ~0.6 & -21\e~8 & & 1.0\e0.1&   ~68\e~4 &   MM2 peak     \\
 (-3.8,~0.0) & ~2.1\e ~0.6 & -10\e~9 & & 0.9\e0.1&   ~68\e~5 &        \\
 (-5.4,~0.0) & ~3.2\e ~1.0 & ~~6\e~9 & & 0.8\e0.2&   ~68\e~6 &        \\
 (-7.0,~0.0) &               &       & & 0.6\e0.2&   ~66\e10 &        \\
 (12.2,~2.0) &               &       & & 1.1\e0.2&   -65\e~6 &        \\
 (10.6,~2.0) &               &       & & 1.0\e0.2&   -70\e~6 &        \\
 (~9.0,~2.0) &               &       & & 0.8\e0.2&   -76\e~7 &        \\
 (~7.4,~2.0) &               &       & & 0.8\e0.2&   -80\e~7 &        \\
 (~5.8,~2.0) &               &       & & 0.7\e0.2&   -85\e~6 &        \\
 (~4.2,~2.0) & ~1.7\e ~0.4 & -57\e~7 & & 0.7\e0.1&   ~89\e~6 & MM 1 peak   \\
 (~2.6,~2.0) &               &       & & 0.7\e0.1&   ~84\e~6 &        \\
 (~1.0,~2.0) &               &       & & 0.8\e0.1&   ~78\e~5 &        \\
 (-0.6,~2.0) &               &       & & 0.9\e0.1&   ~77\e~5 &        \\
 (-2.2,~2.0) &               &       & & 0.9\e0.1&   ~77\e~4 &        \\
 (-3.8,~2.0) &               &       & & 0.9\e0.2&   ~74\e~5 &        \\
 (-5.4,~2.0) &               &       & & 0.8\e0.2&   ~68\e~6 &        \\
 (-7.0,~2.0) &               &       & & 0.6\e0.2&   ~61\e~9 & \\
 (13.8,~4.0) &               &       & & 1.1\e0.3&   -52\e~8 &        \\
 (12.2,~4.0) &               &       & & 1.5\e0.3&   -60\e~5 &        \\
 (10.6,~4.0) &               &       & & 1.4\e0.2&   -65\e~5 &        \\
 (~9.0,~4.0) &               &       & & 1.1\e0.2&   -71\e~5 &        \\
 (~7.4,~4.0) & ~4.3\e ~1.1 & -10\e~7 & & 0.9\e0.2&   -77\e~6 &        \\
 (~5.8,~4.0) & ~1.9\e ~0.6 & -12\e~8 & & 0.8\e0.2&   -86\e~7 &        \\
 (~4.2,~4.0) &               &       & & 0.7\e0.2&   ~87\e~7 &        \\
 (~2.6,~4.0) &               &       & & 0.7\e0.2&   ~85\e~6 &        \\
 (~1.0,~4.0) &               &       & & 0.7\e0.2&   ~83\e~7 &        \\
 (-0.6,~4.0) &               &       & & 0.7\e0.2&   ~84\e~7 &        \\
 (-2.2,~4.0) &               &       & & 0.7\e0.2&   ~85\e~6 &        \\
 (-3.8,~4.0) &               &       & & 0.8\e0.2&   ~81\e~6 &        \\
 (-5.4,~4.0) &               &             & & 0.7\e0.2&   ~75\e~8 &        \\
 (15.4,~6.0) &               &             & & 1.7\e0.6&   -41\e10 &        \\
 (13.8,~6.0) &               &             & & 2.1\e0.4&   -49\e~6 &        \\
 (12.2,~6.0) &               &             & & 2.3\e0.4&   -53\e~4 &        \\
 (10.6,~6.0) &               &             & & 2.1\e0.3&   -55\e~4 &        \\
 (~9.0,~6.0) & ~9.6\e ~3.1 & -~6\e~9 & & 1.6\e0.3&   -60\e~5 &        \\
 (~7.4,~6.0) & ~4.4\e ~1.4 & -10\e~9 & & 1.1\e0.2&   -70\e~6 &        \\
 (~5.8,~6.0) &               &             & & 0.7\e0.2&   -85\e~8 & \\
 (~4.2,~6.0) &               &             & & 0.7\e0.2&   ~88\e~8 &        \\
 (~2.6,~6.0) &               &             & & 0.6\e0.2&   ~87\e~8 &        \\
 (-0.6,~6.0) & ~8.4\e ~2.6 & -93\e~9 & &         &		 & northern region   \\
 (-2.2,~6.0) & 11.7\e ~3.0 & -80\e~7 & & 0.6\e0.2&   -90\e10 & northern region        \\
 (~9.0,~8.0) &               &             & & 2.4\e0.6&   -50\e~7 &        \\
 (~7.4,~8.0) &               &             & & 1.4\e0.4&   -62\e~8 &        \\
 (~5.8,~8.0) &               &             & & 0.8\e0.2&   -76\e~9 &        \\
 (-0.6,~8.0) & 18.0\e ~6.0 & -90\e~9 & &         &		 & northern region   \\
 (-2.2,~8.0) & 22.3\e ~5.8 & -84\e~7 & &         &		 & northern region   \\
 (-2.2,10.0) & 36.4\e 11.2 &-104\e~7 & &           & 		 & northern region   \\

\hline
\hline
\tablenotetext{a}{Offsets are measured with respect to
the phase center: $\alpha_{2000}$=20$^h$39$^m$00\fs72,
$\delta_{2000}$=42\deg22\arcmin46\farcs7.}
\end{tabular}
}
\end{center}
\end{table}

\begin{table}
\begin{center}
\caption{Summary of analysis in polarization angle}
\vspace*{0.3cm}
\begin{tabular}{llcccc}
\hline
\hline
	&  & Average & Observed  & Measurement & Intrinsic \\
Region & Tracer	& Angle	  & Dispersion	& Uncertainty & Dispersion \\
&& $\phi$(\deg)  & $\delta\phi_{obs}$(\deg) &  $\sigma_{\phi}$(\deg) 
& $\delta\phi_{int}$(\deg)\\
\hline
MM1 and MM2 	& Dust 	& -27 & 21\e7 & 7.6\e1.2 & 20\e~7 \\
CO pol Main		& CO 	& ~89 & 23\e5 & 6.3\e1.6 & 22\e~5 \\
\hline
\hline
\end{tabular}
\end{center}
\end{table}

\begin{figure}
\epsscale{0.8}
\plotone{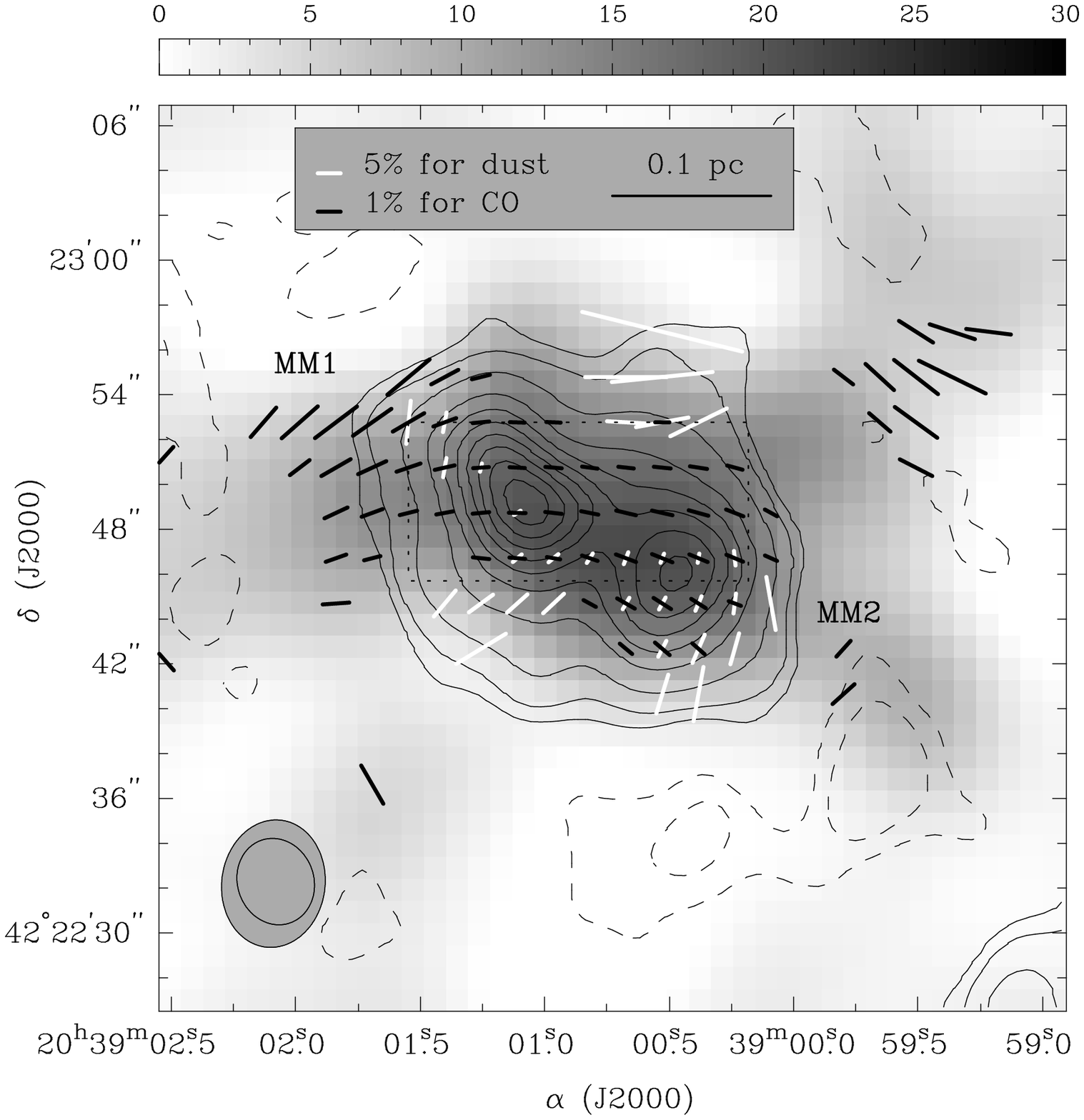}
\caption{Polarization map of DR21(OH). The contours show the Stokes $I$
of the dust continuum
at $-$30, $-$18, 18, 30, 60, 120, 180, 240, 300, 360, 420, and 480 \mJb.
The grey-scale shows the Stokes $I$ of the CO \j21\ emission 
with $v_{LSR}$ of $-$10 \kms, $\Delta v$=11.5 \kms.
The grey-scale bar on the top is in units of \Jb\kms.
The synthesized beam size is 3\farcs9\x3\farcs4 with PA=19\deg\
for the dust continuum and 5\farcs7\x4\farcs6 with PA=$-$7\deg\
for CO, which are shown with grey ellipses.  
The white and black line segments represent the dust and CO
polarization vectors with a scale of 5\% and 1\% per arcsecond
length, respectively.  
The dashed rectangle includes the region associated with
the peak of the polarized CO \j21\ emission,
and the Stokes $I$, $Q$, and $U$ spectra integrated over 
this rectangle are shown in Fig.\ 4. }
\epsscale{1}
\end{figure}

\clearpage

\begin{figure}
\plotone{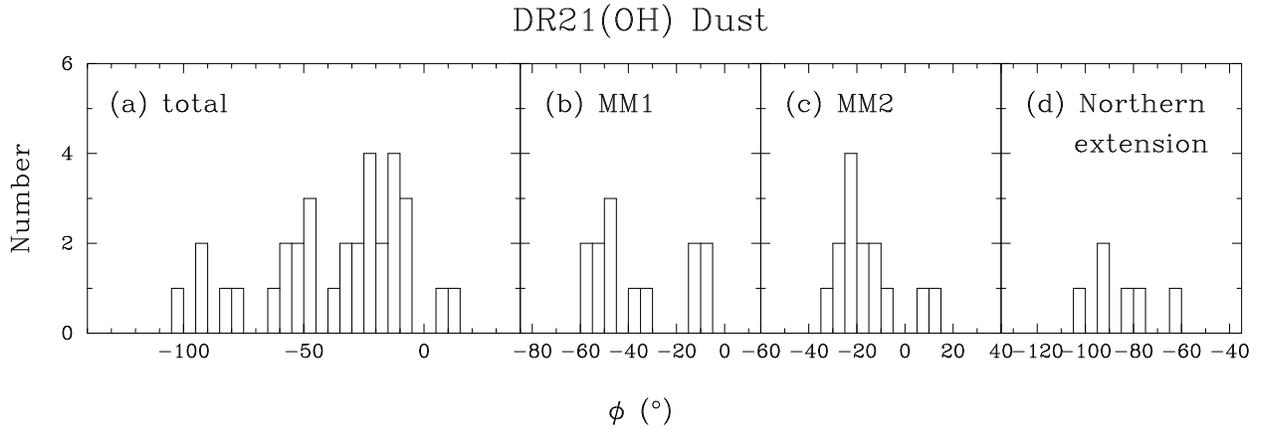}
\caption{Polarization angle distribution of the dust continuum of DR21(OH).
(a) shows the histogram of all dust polarization 
measurements in DR21(OH).  (b)-(d) shows dust polarization
measurements of DR21(OH) in three separate regions.
The vertical axes for these plots are the total number of the
measurements in the bin.
}
\end{figure}

\begin{figure}
\epsscale{0.45}
\plotone{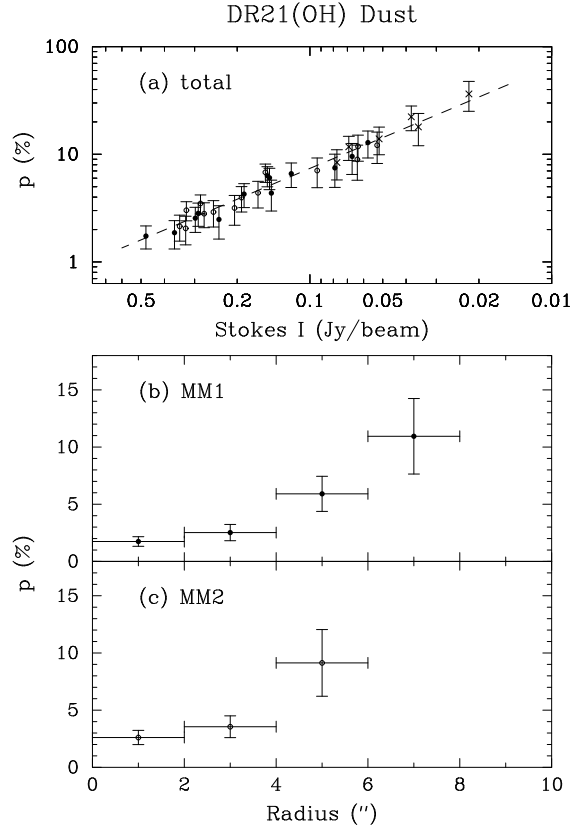}
\epsscale{1}
\caption{Polarization percentage distribution of the dust continuum
of DR21(OH). The filled dots represent the data associated with MM1,
the open dots represent the data associated with MM2, and
the crosses represent the data in the northern edge of the two cores.
The dotted line is the best least-squares fit for the data.
(a) shows the polarization percentage vs. the total intensity.
The error bar of each data point is the measurement error.
The total intensity increases to the left in order to emphasize that
high intensity corresponds to short distance to the center of the cores.
(b) and (c) show the polarization percentage vs. the distance
from MM1 and MM2.  The error bar in radius shows the range over which 
the data are averaged and the error bar in $p$ is the standard deviation 
of the data in the range.}
\end{figure}

\begin{figure}
\epsscale{0.7}
\plotone{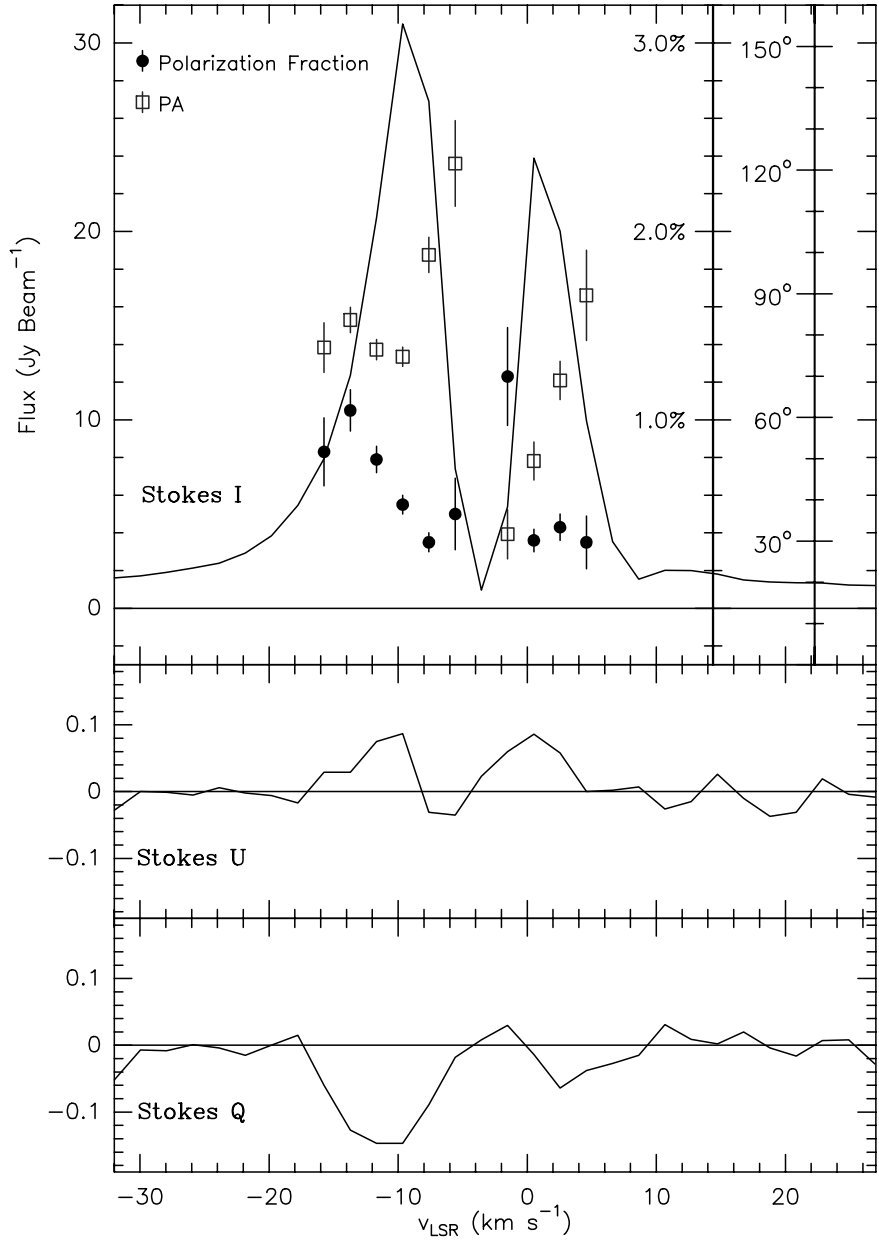}
\epsscale{1}
\caption{CO \j21\ polarization spectra of DR21(OH).
The Stokes $I$, $Q$, and $U$ spectra were integrated over the dashed
rectangle in Fig.\ 1.}
\end{figure}

\begin{figure}
\epsscale{0.5}
\plotone{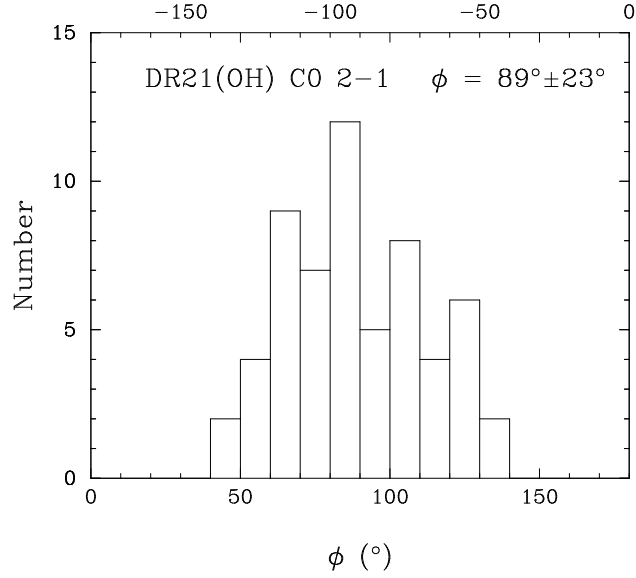}
\epsscale{1}
\caption{Polarization angle distribution of the CO \j21\ emission of
DR21(OH) associated with the dust continuum (CO pol Main).
For easy comparison with the dust polarization,
we label $\phi-180$\deg\ on the top of the plot.
For polarization ``vectors'', $\phi-180$\deg\ is equivalent to $\phi$.
The vertical axis is the total number of the measurements in the bin.
}
\end{figure}

\begin{figure}
\epsscale{0.5}
\plotone{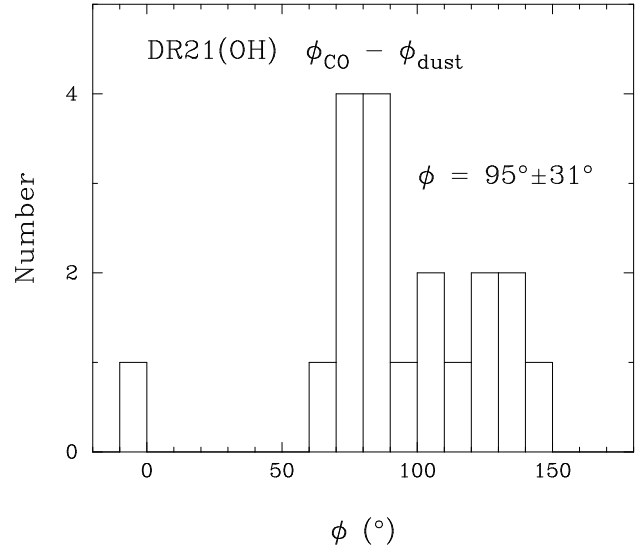}
\epsscale{1}
\caption{Polarization angle distribution of the difference between
CO and dust polarization in DR21(OH).  
The vertical axis is the total number of the measurements in the bin.
The average angle
and dispersion (excluding the data point around 0\deg) is labeled
in the plot. 
}
\end{figure}

\begin{figure}
\epsscale{0.8}
\plotone{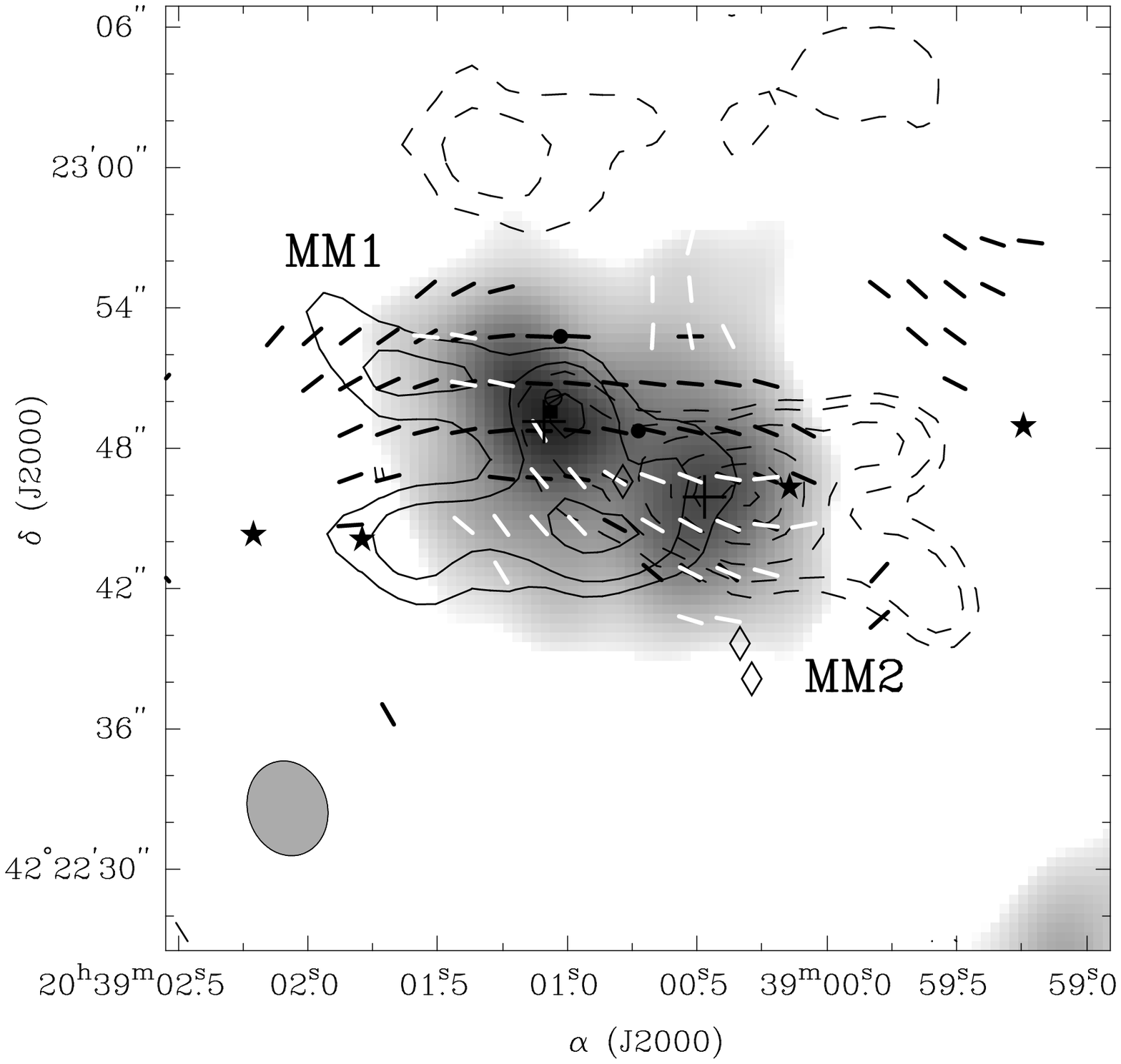}
\epsscale{1}
\caption{The high-velocity outflows in DR21(OH).  The dashed and solid
contours represent the blue and red lobes of CO \j21\ outflows at
the velocity of $-$30 and 20 \kms\ integrated over 7 \kms\ bandwidths.
The beamsize is 4\farcs1\x3\farcs4 with PA=16\deg.
The grey-scale shows the dust continuum at 1.3 mm and
the peaks of MM1 and MM2 are marked by plusses.
The black and white line segments indicate the magnetic field direction
inferred from the CO and dust polarization ( assuming the dust is
aligned by magnetic fields).
The various symbols indicate the maser positions: stars denote
methanol masers (Plambeck \& Menten 1990; filled square represents
the OH maser (Norris et al.\ 1982); filled circles, open circles,
and open diamonds are water masers observed at different epochs
(Genzel \& Downes 1977; Forster et al.\ 1978; Mangum, Wootten, \&
Mundy 1992).  }

\end{figure}

\end{document}